\documentstyle[eqsecnum,aps]{revtex}

\begin{document}
\draft
\preprint{ZIMP-95-08}

\title{$\delta$-function spin-$\frac{1}{2}$ fermions
in a one-dimensional potential well }

\author{You-Quan Li $^{1,2}$ and Jian-Hui Dai $^{2}$ }

\address{
${}^{1}$China Center of Advanced Science and Technology (World Laboratory), P.O. Box 8730 Beijing 100080, China\\
\footnote{mailing address}${}^{2}$Zhejiang Institute of Modern Physics,
Zhejiang University, Hangzhou 310027, China \\
}
\date{Received  11 September, 1995}
\maketitle
\begin{abstract}

The quantum-mechanical problem of $N$ fermions with $\delta$-function
interaction in a one-dimensional potential well of finite depth is
solved. It is shown that there exists exact wave function of
Bethe-ansatz form in the case that a single particle tunnels outside
of the well. The Bethe-ansatz like secular equations for the spectrum
are obtained.

\end{abstract}

\section{Introduction}

One-dimensional quantum-mechanical systems of electrons
have been a subject of great importance in recent years because
they serve as tractable limits of quasi-one-dimensional
systems \cite{Soly} as well as models for the conductive properties
of the one-dimensional quantum wire \cite{Kane}.
A model Hamiltonian describing the system  is a $N$-body
problem with delta-function interaction. For convenience,
let $\hbar$ and the mass of the electron be one.
The Hamiltonian of the electron gas in one dimension then reads
\[
H = - \sum_{i=1}^N \frac{ \partial^2 }{ \partial x_i^2 }
 +  c\sum_{j>i=1}^N \delta( x_{i} - x_{j} )
\]
where $c$ stands for the interaction strength.
The quantum-mechanical problem of the above Hamiltonian
were solved in [4-8] with periodic boundary
conditions, and in \cite{Gaud2} with the boundary condition for
a potential well of infinite depth. Preceding
a very narrow and very deep attractive potential at one side of the
well Ref.\cite{Woyn} considered another kind of boundary condition.
In the discussion of \cite{Woyn}, a phase shift between
the incident and reflected waves, which depends on a phenomenological
parameter, was introduced.  The problem
of many particles with $\delta$-function interaction in a potential well
of finite depth is more complicated due to the fact that the wave
function outside of the well is nonzero and should be taken into account.
In \cite{Li} one of the author solved the  problem of
$N$ bosons with delta-function interaction in a potential well of finite
depth.  The more important physical problem should be
$N$ fermions in a potential well of finite depth. So we will discuss
this problem in the present paper. The Hamiltonian of spin-$1/2$ \,
$N$-particle system interacting by a $\delta$-function in the presence
of a square well is

\begin{equation}
H = - \sum_{i=1}^{N}\frac{\partial^{2} }{\partial x_{i}^{2} }
+ \sum_{i=1}^{N}V(x_{i}) + 2c\sum_{i > j = 1 }^{N}\delta (x_{i} - x_{j} ),
\label{eq:aa}
\end{equation}
where
\[
V(x_{i}) = \left\{
\begin{array}{ll} 
0~~~~~~& |x_i| <{ \displaystyle{ L \over 2 }} \\[6mm]
V_{0}^{2}   & |x_i| > { \displaystyle{ L \over 2 }} \end{array}\right.
\]
in which $L$ is the width of the well. When $V_0=0$, the Hamiltonian is
known to be exactly solvable by the Bethe ansatz in the case of periodic
boundary condition. When $V_0 \neq 0$, the Hamiltonian is not invariant
under translation, and the total momentum of the system is not conserved.
However, the system is still invariant under the action of the
permutation group  $S_N$.
We will diagonalize the Hamiltonian(\ref{eq:aa}) by a generalized
Bethe ansatz which involves all non-diffraction scattering waves.
When the system is considered as a canonical ensemble
in which almost all the particles are confined in the well,
except for one particle which is able to tunnel out of the well,
we will show that the diagonalization can
be exactly realized and the consistent Bethe ansatz equations obtained.

In section \ref{sec:b} we consider the two-particle system to
illustrate the idea of the Bethe ansatz \cite{Bethe}.
In section \ref{sec:c} we analyze the generic system of $N$ fermions and
solve the wave function of Bethe ansatz form.
In section \ref{sec:d}, we diagonalize the $S$-matrix and obtain
secular equations for the spectrum.
In section \ref{sec:e}, We give some discussions and remarks.

\section{Two-particle states}
\label{sec:b}

Before exploring the generalized Bethe ansatz for a system
of $N$ fermions, we discuss the case of two fermions.
The quantum-mechanical problem  of one particle
in a potential well of finite depth is well known.
Because there is no external field, the
spin degrees of freedom $a=+, - $ are degenerate.
The wave functions $\psi_a(x)$ are
\begin{equation}
\psi_a (x)=
\left\{\begin{array}{ll}
A_a^L e^{\kappa x} , & x<-\frac{L}{2} .\\
A_a^+ e^{ikx}+A_a^- e^{-ikx} , & |x|\leq \frac{L}{2} .\\
A_a^R e^{ -\kappa x} , & x>\frac{L}{2} .
\end{array}
\right.
\label{eq:ba}
\end{equation}
where $\kappa>0$ and $k^2=\kappa^2+V_0^2=E$ is the energy of the system.
The amplitudes $A$'s are related via the boundary conditions at
$x=\pm\frac{L}{2}$, which say that the wave function and its derivative
are continuous at the ends of the well.
One can easily find that the spectrum is determined by
$\displaystyle
(\frac{\kappa+ik}{\kappa-ik})^2=e^{-2ikL}
$.

Because of the $\delta$-function interaction  between the two
particles, we should divide the $(x_1,x_2)$ plane into two regions,
namely, the region
${\cal C}(id) = \{ (x_1, x_2) \, | x_1<x_2 \}$
and the region
${\cal C}(\sigma_1) = \{ (x_1, x_2) \,| x_2<x_1 \} $.
The definition of $\sigma_1$ and other terminology will be given
in the next section.
In the square well $x_1, x_2 \in [-\frac{L}{2}, \frac{L}{2} ] $,
the wave function takes the form  of plane waves of all the
non-diffraction scatterings.  We suppose that in the
region ${\cal C}(id)$, the wave function reads
\begin{equation}
\psi_{a_1a_2}^{id}(x_1,x_2) = \sum_{\alpha,\beta =+,-}
[e^{i(\alpha k_1x_1 + \beta k_2x_2)}A_{a_1a_2}(\alpha ,\beta)
- e^{i(\alpha k_1x_2 +\beta k_2x_1)}B_{a_2a_1}(\alpha ,\beta)].
\label{eq:bc}
\end{equation}
The amplitudes $A(\alpha,\beta),B(\alpha,\beta)$ are so chosen
that in the region ${\cal C}(\sigma_1)$, the wave function reads
\begin{equation}
\psi_{a_1a_2}^{\sigma_1}(x_1,x_2)=\sum_{\alpha,\beta =+,-}
[\, e^{i(\alpha k_1x_1+\beta k_2x_2)}B_{a_1a_2}(\alpha ,\beta)
- e^{i(\alpha k_1x_2 +\beta k_2x_1)} A_{a_2a_1}(\alpha ,\beta)\,] .
\label{eq:bd}
\end{equation}
Obviously, the requirement of antisymmetry
for the fermion wave function
$
\psi_{a_1a_2}^{id}(x_1,x_2)=-\psi_{a_2a_1}^{\sigma_1}(x_2,x_1)
$
is satisfied by (\ref{eq:bc}) and (\ref{eq:bd}).

When the interaction strength is not zero,
the wave functions is continuous while its derivative has a finite
difference at the $\delta$-function singularity
\begin{eqnarray}
\psi_{a_1a_2}^{id}(x_1,x_2)|_{x_1=x_2} =
\psi_{a_1a_2}^{\sigma_1}(x_1,x_2)|_{x_1=x_2} , \nonumber  \\
(\frac{\partial}{\partial x_1}-\frac{\partial}{\partial x_2})
(\psi_{a_1a_2}^{\sigma_1}-\psi_{a_1a_2}^{id})(x_1, x_2)|_{x_1=x_2}
 =  2c\psi_{a_1a_2}^{id}(x_1,x_2)|_{x_1=x_2} .
\label{eq:bf}
\end{eqnarray}
Substituting (\ref{eq:bc}) and (\ref{eq:bd}) into (\ref{eq:bf}),
we obtain a relation between the amplitudes $A$ and $B$
\begin{equation}
B_{a_1a_2}(\alpha,\beta)=S_{a_1a_2}^{a_1'a_2'}(\alpha,\beta)
A_{a_1'a_2'}(\alpha,\beta)
\label{eq:bg}
\end{equation}
where
$\displaystyle
S_{a_1a_2}^{a_1'a_2'}(\alpha,\beta)=\frac{
(\alpha k_1-\beta k_2)\delta_{a_1}^{a_1'}\delta_{a_2}^{a_2'}
+2ic\delta_{a_1}^{a_2'}\delta_{a_2}^{a_1'}}
{\alpha k_1-\beta k_2 + 2ic}
$
is the two-body scattering matrix.

When one of the fermions moves out of the potential well, there is an
additional restriction that the wave function should vanish at
infinity. Thus the wave function on the left side of the well takes
the form
$
\psi_{a_1a_2}^L(x_1,x_2)=\sum_{\beta=+,-}
[ e^{\kappa_1x_1}e^{i\beta k_2x_2}A_{a_1a_2}^L(+,\beta)
-e^{\kappa_2x_1}e^{i\beta k_1x_2}B_{a_2a_1}^L(+,\beta)]
$
for $\, x_1<-\frac{L}{2}<x_2<\frac{L}{2}, $
and
$
\psi_{a_1a_2}^L(x_1,x_2)=\sum_{\beta=+,-}
[ e^{\kappa_2x_2}e^{i\beta k_1x_1}B_{a_1a_2}^L(+,\beta)
-e^{\kappa_1x_2}e^{i\beta k_2x_1}A_{a_2a_1}^L(+,\beta) ]
$
for $\, x_2<-\frac{L}{2}<x_1<\frac{L}{2}$.
The former is the antisymmetric counterpart of the later and vice versa.
Similarly, the wave function on the right side of the well should be
$
\psi_{a_1a_2}^R(x_1,x_2)=\sum_{\beta=+,-}
[ e^{-\kappa_2x_2}e^{i\beta k_1x_1}A_{a_1a_2}^R(-,\beta)
-e^{-\kappa_1x_2}e^{i\beta k_2x_1}B_{a_2a_1}^R(-,\beta) ]
$
for
$\, -\frac{L}{2}<x_1<\frac{L}{2}<x_2,$
and
$\psi_{a_1a_2}^R(x_1,x_2)=\sum_{\beta=+,-}
[ e^{-\kappa_1x_1}e^{i\beta k_2x_2}B_{a_1a_2}^R(-,\beta)
-e^{-\kappa_2x_1}e^{i\beta k_1x_2}A_{a_2a_1}^R(-,\beta) ]
$
for
$\, -\frac{L}{2}<x_2<\frac{L}{2}<x_1$.

The amplitudes $A^{R,L}(\alpha,\beta), B^{R,L}(\alpha,\beta)$
are related to the amplitudes $A(\alpha,\beta), B(\alpha.\beta)$
via the boundary conditions at
$ x=\pm \frac{L}{2}$, namely
\begin{eqnarray*}
\psi_{a_1a_2}^L(-\frac{L}{2},x_2) & = &
\psi_{a_1a_2}(-\frac{L}{2},x_2),\nonumber \\
\frac{\partial}{\partial x_1}\psi_{a_1a_2}^L(x_1,x_2)|_{x_1=- L/2}
& = &
\frac{\partial}{\partial x_1}\psi_{a_1a_2}(x_1,x_2)|_{x_1=- L/2}
\end{eqnarray*}
and
\begin{eqnarray*}
\psi_{a_1a_2}^R(x_1,\frac{L}{2}) & = & \psi_{a_1a_2}(x_1,\frac{L}{2}) ,\\
\frac{\partial}{\partial x_2}\psi_{a_1a_2}^R(x_1,x_2)|_{x_2= L/2}
& = &
\frac{\partial}{\partial x_2}\psi_{a_1a_2}(x_1,x_2)|_{x_2= L/2} .
\end{eqnarray*}
Thus we get
\begin{equation}
e^{-ik_1L}\frac{\kappa_1-ik_1}{\kappa_1+ik_1}=
-\frac{A_{a_1a_2}(-,\beta)}{A_{a_1a_2}(+,\beta)} ,
\label{eq:bo}
\end{equation}
\begin{equation}
e^{-ik_2L}\frac{\kappa_2-ik_2}{\kappa_2+ik_2}=
-\frac{B_{a_2a_1}(\beta, -)}{B_{a_2a_1}(\beta, +)} ,
\label{eq:bp}
\end{equation}
\begin{equation}
e^{-ik_1L}\frac{\kappa_1-ik_1}{\kappa_1+ik_1}=
-\frac{B_{a_2a_1}(+,\beta)}{B_{a_2a_1}(-,\beta)} ,
\label{eq:bq}
\end{equation}
and
\begin{equation}
e^{-ik_2L}\frac{\kappa_2-ik_2}{\kappa_2+ik_2}=
-\frac{A_{a_1a_2}(\beta, +)}{A_{a_1a_2}(\beta, -)} ,
\label{eq:br}
\end{equation}
where $\beta = +, - $. We observe that the dependence of
$A^{R,L}, B^{R,L}$ on $A,B$
are the same as what occurred in the one particle situation.

The equations (\ref{eq:bo}-\ref{eq:br}) define the secular equations
for the spectrum of quasi momentum ( charge ) $k_1,k_2$.
This is due to the fact that
\begin{equation}
S^{-1}(k_1,k_2)S(k_1,-k_2)=-S^{-1}(-k_2,k_1)S(k_2,k_1).
\label{eq:bs}
\end{equation}
The secular equation for the spectrum is
\begin{eqnarray}
S^{-1}(-,\beta)S(+,\beta)A(-,\beta)=f_1^2A(-,\beta) , \nonumber \\
S^{-1}(\beta,+)S(\beta,-)A(\beta, +)=f_2^2A(\beta, +) , \, \beta=+,-.
\label{eq:bt}
\end{eqnarray}
where
$
f_j = -\displaystyle\frac{\kappa_j-ik_j}{\kappa_j+ik_j}e^{-ik_jL},
\, j=1,2.
$
After diagonalizing the scattering operators $S$ and $S^{-1}$,
one can obtain the
Bethe ansatz equations for $k_1,k_2$.
This will be discussed in section \ref{sec:d}.

\section{The case of  N fermions }
\label{sec:c}
                                                                    
We first introduce some terminology which is helpful for avoiding
the ambiguities which have appeared in some previous literature.
In a Euclidean space ${\sf l \! R }^{N}$ with Cartesian coordinates
$ x = (x_1 , x_2 , \cdots , x_N ),$
the set of hyperplanes
$ \{ \{ x | x_i - x_j = 0 \} |~ i, j = 1, 2, \cdots N \}$
partition 
${\sf l \! R }^{N} $
into finitely many regions. We use the convention that the scalar
product of two $N$-dimensional vectors is written as
$ ( x | y) = \sum^{N} _{i=1}(x)_i (y)_i $.
The above-mentioned hyperplanes are Weyl reflection hyperplanes
$ ( x| \alpha_{ij} ) = 0 $
where
$ \alpha_{ij} = e_i -e_j $
are the roots of the Lie algebra $ A_{N-1} $.
We denote a Weyl reflection hyperplane as
$ { \sf l \! P}_{\alpha} := \{ x ~| ~~( x| \alpha )=0 \} $, in which
$\alpha $  is a root of $ A_{N-1} $.
The connected components of 
$ {\sf l \! R}^{N} \setminus \{ {\sf l \! P }_{\alpha} \} $
are called Weyl Chambers \cite{Gilmore}
of the Lie algebra $ A_{N-1} $. We denote
the Weyl group of $ A_{N-1} $ as $ {\cal W}_A $, the basic elements of
which are defined by
$
\sigma_{i} : ( x_{1}, x_{2}, \cdots , x_{i}, x_{i+1}, \cdots , x_{N}~)
\mapsto ( x_{1}, x_{2}, \cdots , x_{i+1}, x_{i}, \cdots , x_{N}~).
$
The Weyl group of $B_N$, denoted by ${\cal W}_B$, has one more basic
element $\bar{\sigma}_N$ besides the previously defined
$\sigma_i (i=1, 2, \cdots N-1) $. The definition of $\bar{\sigma}_N$
is given by
$
\bar{\sigma}_N : ( x_1, \cdots, x_{N-1}, x_N ) \mapsto
( x_1, \cdots, x_{N-1}, - x_N )
$.
An element of the Weyl group obviously maps one Weyl chamber onto another
and all the chambers can be obtained from any given chamber via the
actions of the whole Weyl group. 
Thus Weyl chambers can be specified by the elements of Weyl group.

Now we consider the case of $N$ fermions.
Obviously, the Hamiltonian (\ref{eq:aa}) is invariant under the
action of the permutation group $S_N $, but it is not invariant under
translation. Thus the total momentum of the system is not conserved.
The Schr\"{o}dinger equation of the Hamiltonian (\ref{eq:aa}) on
the domain 
$ {\sf l \! R}^{N} {\sf  \setminus \{ l \! P }_{\alpha} \}  $
becomes
\[
\sum_{i=1}^{N}
\left[ - \frac{\partial^{2} }{\partial x_{i}^{2} }
+ V(x_{i})
\right] \psi_a (x) = E \psi_a (x),
\]
where $ a:= (a_1, a_2, \cdots, a_N )$ and
$ a_i \in \{ +, - \}$. Solutions of the equation are plane waves.
Considering that the total momentum is not conserved, we adopt
the following Bethe\cite{Bethe} ansatz  form

\begin{equation}
\psi_a (x)=\sum_{\sigma\in W_B}A_a(\sigma,\tau)e^{i(\sigma k|x)},
\, x \in {\cal C}(\tau)
\label{eq:ca}
\end{equation}
where
$\sigma k $
stands for the image of a given 
$ k:= ( k_1 , k_2, \cdots , k_N ) $
by a mapping 
$\sigma \in {\cal W}_{B} $ 
and the coefficients 
$ A(\sigma, \tau )$ 
are functionals on 
${\cal W}_{B} \otimes {\cal W}_{A}$.
We emphasize that the sum runs over the Weyl group of the Lie algebra
$B_N $ but the wave function is defined on various Weyl chambers
corresponding to the Weyl group of the Lie algebra $A_{N-1}$.
This is different from the situation of periodic boundary condition.

For a fermionic system, the wave function is supposed to be
anti-symmetric under any permutation of  both  coordinates and
spin states, i.e.

\begin{equation}
(\sigma_i\psi)_a(x)=-\psi_a(x).
\label{eq:cb}
\end{equation}
Here $(\sigma_i \psi)_a$ is well defined by
$\psi_{\sigma_i a}(\sigma^{-1}_i x)$.
Therefore both sides of (\ref{eq:cb}) can be written out by
using (\ref{eq:ca}).
Furthermore, using the evident identity
$ (\sigma k | \sigma_{i}^{-1} x ) = (\sigma_{i} \sigma k | x ) $
and the rearrangement theorem of group theory, we obtain the 
following consequence from (\ref{eq:cb})
\begin{equation}
A_a(\sigma,\sigma_i\tau)=-A_{\sigma_ia}(\sigma_i\sigma,\tau).
\label{eq:cf}
\end{equation}

The $\delta$-function term in the Hamiltonian (\ref{eq:aa})
contributes a boundary condition
at hyperplane $ {\sf l \! P }_{\alpha} $ ($\alpha$ is a root of Lie
algebra $A_{N-1}$), namely a discontinuity
of the derivative of
wave function along the normal of Weyl hyperplane:
\begin{eqnarray}
\lim_{\epsilon\rightarrow 0^+}
[\alpha\cdot\nabla\psi_a(x_{(\alpha)}+\epsilon\alpha)
-\alpha\cdot\nabla\psi_a(x_{(\alpha)}-\epsilon\alpha)]
& = & 2c\psi_a(x_{(\alpha)}),    \nonumber \\
\lim_{\epsilon\rightarrow 0^+}
[\psi_a(x_{(\alpha)}+\epsilon\alpha)-\psi_a(x_{(\alpha)}-\epsilon\alpha)]
& = & 0,
\label{eq:cd}
\end{eqnarray}
where $x_{(\alpha)} \in {\sf l \! P }_{\alpha} $
and $ \nabla := \sum^{N}_{i=1} e_i (\partial / \partial x_i )$.

Substituting (\ref{eq:ca}) into (\ref{eq:cd}), we find that
\begin{equation}
i[(\sigma k)_i-(\sigma k)_{i+1}]
[A_a(\sigma,\sigma_i\tau)-A_a(\sigma_i\sigma,\sigma_i\tau)
-A_a(\sigma,\tau)+A_a(\sigma_i\sigma,\tau)]
=2c [A_a(\sigma,\tau)+A_a(\sigma_i\sigma,\tau].
\label{eq:ce}
\end{equation}
By making use of the relation (\ref{eq:cf}),
we can obtain the following relations
from (\ref{eq:ce})
\begin{equation}
A_a(\sigma_i\sigma,\tau)=S_{a,a'}^i(\sigma k)A_{a'}(\sigma,\tau).
\label{eq:cg}
\end{equation}
\begin{equation}
S_{a,a'}^i(\sigma k)= -
\frac{c \delta_{a, a'}-i[(\sigma k)_i-(\sigma k)_{i+1}\,] P_{a,a'}}
     {c-i[(\sigma k)_i-(\sigma k)_{i+1}] \,\,  } .
\label{eq:ch}
\end{equation}
where $ a'= \sigma_i a$ and $P_{a, a'}$
stands for the matrix elements of
the spinor representation of permutation group.

The relation (\ref{eq:cf}) provides  for the coefficients
A a relation between different Weyl chambers.
Eq.(\ref{eq:cg}) provides a connection between those coefficients which
are related via any element of Weyl group ${ \cal W}_{A} $ in the same
Weyl chamber. Although the $ 2^{N} N! $ coefficients are determined by 
(\ref{eq:cg}) only  up to $ 2^{N}$ arbitrary constants, we will see
in next section that the secular equation for the spectrum is determined
uniquely. The basic elements of the Weyl group ${ \cal W}_{A} $ obey
$ \sigma_{i}^{2} = 1 ~~{\rm and }~~ 
\sigma_{i} \sigma_{i+1} \sigma_{i}=
\sigma_{i+1} \sigma_{i} \sigma_{i+1}$ 
as identities. These identities must involve some relations i.e.
$
A_a(\sigma_{i}^{2} \sigma, \tau) = A_a(\sigma, \tau)
$ and
$
 A_a(\sigma_{i}\sigma_{i+1}\sigma_{i}\sigma, \tau)=
 A_a(\sigma_{i+1}\sigma_{i}\sigma_{i+1}\sigma, \tau)
$.
Using (\ref{eq:cg}) repeatedly, one can obtain the following relations:
\begin{eqnarray}
S^i (\sigma_i \sigma k)S^i (\sigma k) & = & I,\nonumber \\
S^i (\sigma_{i+1}\sigma_i \sigma k)
S^{i+1}(\sigma_i\sigma k)
S^i (\sigma k) & = &
S^{i+1}(\sigma_i \sigma_{i+1}\sigma k)
S^i (\sigma_{i+1}\sigma k)
S^{i+1}(\sigma k),
\label{eq:ci}
\end{eqnarray}
where we have adopted the conventions $ S = matrix(S_{ab})$,
$S^i = S\otimes I, \, S^{i+1}= I \otimes S \,$
( $I$ is a $2\times 2$ unit matrix).
These relations are consistency conditions for the S-matrix.
The second relation is called a Yang-Baxter equation.
The concrete S-matrix in (\ref{eq:ch}) verifies these relations.

\section{ Diagonalization of the S-matrix and the
secular equation for the spectrum }
\label{sec:d}

Because the antisymmetric property provides a relation for wave
function between different Weyl chambers, we only need to discuss
the problem in one chamber.
For simplicity we consider ${ \cal C}(id)$ i.e. the region
$ 
x_1 < x_2 < \cdots < x_N 
$.
Owning to the potential well terms in Hamiltonian,
we must consider the wave function in the three different regions
${\displaystyle
x_{1} < -{L \over 2 } < x_{2} < \cdots < x_{N} < {L \over 2 },
}$
${\displaystyle
 -{L \over 2 }< x_{1} < x_{2 } < \cdots < x_{N } < {L \over 2 }
} $ and 
${ \displaystyle
 -{L \over 2 }< x_{1} < x_{2} < \cdots < x_{N-1} < {L \over 2 } < x_{N}
}$
respectively. The cases of more than one particle being outside of the
interval 
${\displaystyle
[-{L \over 2} , {L \over 2 } ]  }
$
are not important since those regions are not next to the region
$
\displaystyle
-{L \over 2 }< x_1 < x_2 < \cdots < x_N < {L \over 2 }
$.

We first make some notation conventions:
$
{\cal W}':=\{\sigma_2 , \sigma_3 , \cdots,\sigma_{N-1},\bar{\sigma}_N \,\}
$
and
$
{\cal W}'':=\{\sigma_1 , \sigma_2 , \cdots, \sigma_{N-2},
\sigma_{N-1}\bar{\sigma}_N \sigma_{N-1} \}
$
are two subgroups of the Weyl group of $B_N $;
two particular cycles are
$ 
q'_{j}: (x_1 , \cdots , x_{j}, \cdots , x_N ) \mapsto 
(x_j , x_1 , \cdots , x_{j-1}, x_{j+1}, \cdots , x_{N})
$
and
$ q''_{j}: (x_1 , \cdots , x_{j}, \cdots , x_N ) \mapsto
( x_1 , \cdots , x_{j-1}, x_{j+1}, \cdots , x_{N}, x_j ) 
$.
Then we can write (\ref{eq:ca}) into a sum of pairs and get
the following expressions
( because there will be no confusion here, we adopt
the notation $\psi_a $ instead of $ \psi^{id}_a $ ):

\noindent (i) In the region
$ {\displaystyle
-{L \over 2 } < x_1 < x_2< \cdots < x_N < { L \over 2 }  
 }$,
\begin{eqnarray}
\psi_a (x) & = & \sum_{j=1}^N \sum_{\sigma' \in {\cal W}' }
[ \,A_a (\sigma' q'_{j}, id ) e^{i k_{j} x_{1} }
+ A_a (\bar{\sigma}_{1}\sigma' q'_{j}, id ) e^{ -i k_{j} x_{1} }\,]
e^{ i (\sigma' q'_{j} k | x' ) } \nonumber \\
\, & \stackrel{ \rm or }{ = } & \sum_{j=1}^{N}\sum_{\sigma''\in{\cal W}''}
[ \,A_a (\sigma'' q''_{j}, id ) e^{i k_{j} x_{N} }
+ A_a (\bar{\sigma}_{N}\sigma'' q''_{j}, id ) e^{ -i k_{j} x_{N} } \,]
e^{ i (\sigma'' q''_{j} k | x'' ) }
\label{eq:da}
\end{eqnarray}
here and the following
$ 
x':= ( 0, x_2 , \cdots , x_N ),~~
 x'':= ( x_1 , x_2 , \cdots , x_{N-1}, 0 ) ~
$
and  
$
\bar{\sigma}_{1}:= \sigma_{1}\sigma_{2}\cdots \sigma_{N-1}
\bar{\sigma}_{N}\sigma_{N-1}\cdots \sigma_{2}\sigma_{1};
$

\noindent (ii) In the region
${\displaystyle
x_1 < -{L \over 2 }< x_2 < \cdots < x_N < {L \over 2 }
} $,
\begin{equation}
\psi_a^{L} (x) = \sum_{j=1}^{N}\sum_{\sigma' \in {\cal W}' }
A^{L}_a (\sigma' q'_{j}, id ) e^{ \kappa_{j} x_{1} }
e^{ i (\sigma' q'_{j} k | x' ) }
\label{eq:db}
\end{equation}
here  and the following
$ 
\kappa_{j} > 0 , ~~ \kappa_{j}^{2} + k_{j}^{2} = V_{0}^{2}
$;

\noindent (iii) In the region
${\displaystyle
 -{L \over 2 }< x_{1} < x_{2} < \cdots < x_{N-1} <
 {L \over 2 }< x_{N} }$,
\begin{equation}
\psi_a^{R}(x) = \sum_{j=1}^{N}\sum_{\sigma'' \in {\cal W}'' }
A^{R}_a (\sigma'' q''_{j}, id ) e^{ - \kappa_{j} x_{N} }
e^{ i (\sigma'' q''_{j} k | x'' ) }.
\label{eq:dc}
\end{equation}
Because of the finite depth of potential well,
both the wave function and
its derivative are continuous at the ends of the well.
At the left end, we have
\begin{eqnarray}
\psi_a^{L} \vert _{ x_{1}  = - L/2 }
& = & \psi_a \vert _{x_{1} = - L/2  }\nonumber \\
\frac{\partial }{\partial x_{1} }\psi_a^{L} \vert _{x_1=- L/2}
& = & \frac{\partial}{\partial x_{1} }\psi_a \vert_{x_1=-L/2}
\label{eq:de}
\end{eqnarray}
Thus we get
$A_a^L(\sigma'q_j',id)=A_a(\sigma'q_j',id)e^{-(ik_j-\kappa_j)L/2}
+ A_a(\bar{\sigma}_1\sigma'q_j',id)e^{(ik_j+\kappa_j)L/2}$  and
\begin{equation}
\frac{\kappa_{j} - i k_{j}}{\kappa_{j} + i k_{j} }
= -\, \frac{ A_a( \bar{\sigma }_{1}\sigma' q'_{j}, id )}
{ A_a(\sigma' q'_{j}, id ) }\,e^{ i k_{j} L }.
\label{eq:df}
\end{equation}
Similarly, the conditions at the right end read
\begin{eqnarray}
\psi_a^{R} \vert _{ x_{2}  = L/2 }
& = & \psi_ a \vert _{x_{2} = L/2 }\nonumber \\
\frac{\partial }{\partial x_{2} }\psi_a^{R} \vert _{x_{2}=L/2 }
 & = & \frac{\partial }{\partial x_{2} }\psi_a \vert_{x_{2}=L/2}
\label{eq:dg}
\end{eqnarray}
They give rise to
$A_a^R(\sigma''q_j'',id)=A_a(\sigma''q_j'',id)e^{(ik_j+\kappa_j)L/2}
+A_a(\bar{\sigma}_N\sigma''q_j'',id)e^{(-ik_j+\kappa_j)L/2}$  and
\begin{equation}
\frac{\kappa_{j} + i k_{j}}{\kappa_{j} - i k_{j} }
= - \,\frac{ A_a( \bar{\sigma }_{N}\sigma'' q''_{j}, id )}
{ A_a(\sigma'' q''_{j}, id ) } \, e^{ - i k_{j} L }.
\label{eq:dh}
\end{equation}

By making use of (\ref{eq:da}) and (\ref{eq:cg}),
we obtain the matrix relations
\begin{eqnarray}
A(\sigma q_j',id)=S_1(\sigma_2\cdots\sigma_{N-1}\sigma q_j''k)
                   S_2(\sigma_3\cdots\sigma_{N-1}\sigma q_j''k)\nonumber \\
        \cdots     S_{N-2}(\sigma_{N-1}\sigma q_j''k)
                   S_{N-1}(\sigma q_j''k)
        \cdot      A(\sigma q_j'',id)
\label{eq:di}
\end{eqnarray}
\begin{eqnarray}
A(\bar{\sigma}_1\sigma q_j',id)
     = S_1(\sigma_2\cdots\sigma_{N-1}\bar{\sigma}_N\sigma q_j''k)
     S_2(\sigma_3\cdots\sigma_{N-1}\bar{\sigma}_N\sigma q_j''k)\nonumber\\
\cdots
     S_{N-2}(\sigma_{N-1}\bar{\sigma}_N\sigma q_j''k)
     S_{N-1}(\bar{\sigma}_N\sigma q_j''k)
\cdot
     A(\bar{\sigma}_N\sigma q_j'',id).
\label{eq:dj}
\end{eqnarray}
Here and in the following equations the spin labels are omitted,
hence they should be understood as matrix equations.
From (\ref{eq:df},\ref{eq:dh}) together with
(\ref{eq:di},\ref{eq:dj}) we obtain an
eigenequation in the spin space:
\begin{equation}
\bar{T}_j^{-1}\bar{T}_j A(\bar{\sigma}_N\sigma q_j'',id)=f_j^2
                        A(\bar{\sigma}_N\sigma q_j'',id)
\label{eq:dl}
\end{equation}
where
\begin{equation}
T_j=S^{jj-1}\cdots S^{j1}S^{jN}\cdots S^{jj+1}, \,
\bar{T}_j =
\bar{S}^{jj-1}\cdots\bar{S}^{j1}\bar{S}^{jN}\cdots\bar{S}^{jj+1}
\label{eq:dm}
\end{equation}
and
\begin{equation}
S^{ji}\equiv S_i(\sigma_{i+1}\cdots\sigma_{N-1}\sigma q_j''k)
=-\displaystyle{\frac{[(\sigma k)_j-(\sigma k)_i]P-ic}
               {(\sigma k)_j-(\sigma k)_i+ic}}
\label{eq:dn}
\end{equation}
\begin{equation}
\bar{S}^{ji}\equiv S_i(\sigma_{i+1}\cdots\sigma_{N-1}\bar{\sigma}_N
\sigma q_j''k)
=-\displaystyle{\frac{[-(\sigma k)_j-(\sigma k)_i]P-ic}
               {-(\sigma k)_j-(\sigma k)_i+ic}}.
\label{eq:do}
\end{equation}

In order to solve the eigenequation (\ref{eq:dl}),
we observe the highest weight state corresponding to $M$ down-spins
and $N - M$ up-spins,
\begin{equation}
\Phi^M = \sum_{ \{ m_{\mu} \} }
\varphi_{ \{ m_{\mu } \} }
\prod_{\mu = 1 }^{M} \hat{s}_{ m_{\mu} }^{-} \phi_{++ \cdots +}
\label{eq:dwf}
\end{equation}
where
$ \phi_{++\cdots +}$ stands for the state of $N$ up-spins,
$\hat{s}^-_{m_{\mu} }$ flips a up-spin to a down-spin at the
$m_{\mu}$th position.
The state (\ref{eq:dwf} ) is checked to be a common eigenstate of both
operators $T_j $ and $\bar{T}_j $ if $\varphi $ takes the following form
\begin{equation}
\varphi_{ m_1 m_2 \cdots m_M } =\sum_{\tilde{\sigma}}
\Gamma(\tilde{\sigma})\prod^M_{\mu=1}
F(\,(\tilde{\sigma}\lambda )_{\mu}, m_{\mu})
\label{eq:dp}
\end{equation}
where
\begin{equation}
\frac{\Gamma(\tilde{\sigma}_i\tilde{\sigma}\lambda)}
{\Gamma(\tilde{\sigma}\lambda)}=
\frac{[i(\tilde{\sigma} \lambda)_{i+1}-(\tilde{\sigma} \lambda)_i-c]}
     {[i(\tilde{\sigma} \lambda)_{i+1}-(\tilde{\sigma}\lambda)_i+c]},\,\,
\frac{\Gamma(\bar{\sigma}_N\tilde{\sigma}\lambda)}
{\Gamma(\tilde{\sigma}\lambda)}=-1.
\label{eq:dq}
\end{equation}
and
\begin{equation}
F(\lambda;m)= \frac{1}{ i(k_m-\lambda)+ c/2 }
 \prod_{j=1}^{m-1}
  \frac{i(k_j-\lambda)- c/2 }
       {i(k_j-\lambda)+ c/2 },
\label{eq:dr}
\end{equation}
where $\tilde{\sigma}$ stands for the elements of the Weyl group of the Lie
algebra $B_M$; $\tilde{\sigma}_i \, ( i= 1,2,\cdots, M ) $ and
$ \bar{\sigma}_M $ stands  for its basic elements.
The parameter $\lambda_{\mu} \, (\mu = 1, 2, \cdots, M ) $
for the spin wave function is supposed to satisfy
\begin{equation}
\prod_{\nu\neq\mu}^M
\frac{\lambda_{\mu}-\lambda_{\nu}+ic}
     {\lambda_{\mu}-\lambda_{\nu}-ic} \cdot
\frac{\lambda_{\mu}+\lambda_{\nu}+ic}
     {\lambda_{\mu}+\lambda_{\nu}-ic}
=\prod_{j=1}^N
\frac{\lambda_{\mu}-k_j-ic/2}
      {\lambda_{\mu}-k_j+ic/2} \cdot
\frac{\lambda_{\mu}+k_j-ic/2}
     {\lambda_{\mu}+k_j+ic/2}
\label{eq:dt}
\end{equation}
This is a generalization of Bethe-Yang ansatz\cite{Yang,Woyn}.
The eigenvalues corresponding to the operators $T_j$ and
$\bar{T}^{-1}_j$ are

\begin{equation}
z_j=\prod_{\nu=1}^M\frac{\lambda_{\nu}-k_j-ic/2}
                          {\lambda_{\nu}-k_j+ic/2}, \,\,
\bar{z}_j^{-1}=\prod_{\nu=1}^M\frac{\lambda_{\nu}+k_j-ic/2}
                                     {\lambda_{\nu}+k_j+ic/2}
\label{eq:ds}
\end{equation}
respectively. Hence we get the coupled Bethe ansatz equations for the
variables $\lambda_{\mu}, (\mu=1,2,\cdots,M).$:
and the quasi-momenta $z_j\bar{z_j}^{-1}=f_j^2$, explicitly
\begin{equation}
\frac{\kappa_j-ik_j}{\kappa_j+ik_j}^2e^{-2ik_jL}=
\prod_{\nu=1}^M\frac{\lambda_{\nu}-k_j-ic/2}
                       {\lambda_{\nu}-k_j+ic/2}\cdot
\frac{\lambda_{\nu}+k_j-ic/2}
                       {\lambda_{\nu}+k_j+ic/2}.
\label{eq:du}
\end{equation}

Taking the logarithm of (\ref{eq:du}) and (\ref{eq:dt}) we obtain
the secular equations
\begin{equation}
k_j=\frac{\pi}{L}I_j-\frac{1}{L}\sin^{-1}(\frac{k_j}{V_0})
-\frac{1}{L}\sum_{\nu=1}^M
\left[
     \tan^{-1}(\frac{k_j-\lambda_{\nu}}{c/2})
     +\tan^{-1}(\frac{k_j+\lambda_{\nu}}{c/2})
\right]
\label{eq:dv}
\end{equation}

\begin{equation}
\sum_{j=1}^N
\left[
     \tan^{-1}(\frac{\lambda_{\mu}-k_j}{c/2})
     +\tan^{-1}(\frac{\lambda_{\mu}+k_j}{c/2})
\right]
=\pi J_{\mu}
+ \sum_{\nu\neq\mu}^M
\left[ \tan^{-1}(\frac{\lambda_{\mu}-\lambda_{\nu}}{c})
      +\tan^{-1}(\frac{\lambda_{\mu}+\lambda_{\nu}}{c})
\right].
\label{eq:dw}
\end{equation}
where both $I_j$ and $J_{\mu}$ are integers, that play the
roles of quantum numbers. These equations determine
the spectrum $\{k_j\}$ of the
system with total spin $S_z=\frac{1}{2}N-M$ and the total energy
$E=\sum_{j=1}^Nk_j^2$.

\section{Conclusions and remarks}
\label{sec:e}

In the previous sections, we solved the problem of $N$ fermions with
$\delta$-function interaction in a one dimensional potential well of
finite depth. In our discussion, we considered the special case in
which a single particle tunnels out of the potential well. In this case,
we showed that there exists a Bethe ansatz like exact solution
for the problem.
We also obtained the secular equations which determine the spectrum
of $N$-body system. Differing from the usual $\delta$-function problem,
there is a contribution to the secular equations by the reflection of the
particles at the ends of the potential well. This contribution affects both
the charge and spin excitations. A new feature of present model is the
additional term proportional to
$\sin^{-1}(k_j/V_0)$ in (\ref{eq:dv}). As a phase shift of one particle
at the barrier of potential well of finite depth, this term  contributes
to charge excitation only. The phase shift vanishes when
$V_0 \rightarrow \infty $, which agrees with the result for the
potential well of infinite depth. In the thermodynamic limit
($L \rightarrow \infty, \, N \rightarrow \infty \,
{\rm and}\, M \rightarrow \infty $ but $N/L-M/L$ keeps a finite value),
equations (\ref{eq:dv}) and (\ref{eq:dw}) will become integral equations.
Certainly, one can discuss the ground state
properties and its thermodynamics in the spirit of \cite{YY}.

We known that the $N$ particles were thought of as
situated on a circle in the case of periodic boundary conditions.
For the  problem of $N$ electrons moving in a quantum wire of
finite length, the periodic boundary condition is no longer
available. Then we may consider the boundary as a potential well.
If all the $N$ electrons are confined in an interval, the potential
well is supposed to have infinite depth.
Strictly speaking, electrons are able to leave the ends of
the wire due to the tunneling effect.
Then the more appropriate boundary condition is a potential well
of finite depth. Different from  the case of
infinite depth in which one needs only to consider the wave function
in the potential well, we should take into account the possibility
of a non-vanishing wave function outside of the well in the case of
finite depth.
The single particle tunneling process is the simplest non-trivial
one for the problem of the potential well of finite depth.
The result of present paper is the first step toward
the exact solution of this problem.

This work is supported by NNSF of China and NSF of Zhejiang province.

\end{document}